\documentclass[lettersize,journal]{IEEEtran}
\usepackage{amsmath,amsfonts}
\usepackage{algorithmic}
\usepackage{algorithm}
\usepackage{array}
\usepackage[caption=false,font=normalsize,labelfont=sf,textfont=sf]{subfig}
\usepackage{textcomp}
\usepackage{stfloats}
\usepackage{url}
\usepackage{verbatim}
\usepackage{graphicx}
\usepackage{cite}
\begin{document}

\title{A Novel Hybrid Digital and Analog Laser Synchronization System}

\author{Mingwen Zhu, Shangsu Ding,~\IEEEmembership{Graduate Student Member,~IEEE}, Tianwei Jiang,~\IEEEmembership{Member,~IEEE}, Jianming Shang, Song Yu, Bin Luo,~\IEEEmembership{Member,~IEEE},
        % <-this % stops a space
\thanks{This paper was produced by the IEEE Publication Technology Group. They are in Piscataway, NJ.}% <-this % stops a space
\thanks{Manuscript received April 19, 2021; revised August 16, 2021.}}

\markboth{Journal of \LaTeX\ Class Files,~Vol.~14, No.~8, August~2021}%
{Shell \MakeLowercase{\textit{et al.}}: A Sample Article Using IEEEtran.cls for IEEE Journals}


\maketitle

\begin{abstract}
Laser synchronization is a technique that locks the wavelength of a free-running laser to that of the reference laser, thereby enabling synchronous changes in the wavelengths of the two lasers. This technique is of crucial importance in both scientific and industrial applications. Conventional synchronization systems, whether digital or analog, have intrinsic limitations in terms of accuracy or bandwidth. The hybrid "digital + analog" system can address this shortcoming. However, all above systems face the challenge of achieving an both high locking accuracy and low structural complexity simultaneously. This paper presents a hybrid "digital + analog" laser synchronization system with low-complexity and high-performance. In the digital part, we proposed a electric intensity locking method based on a band-pass filter, which realizes the fluctuation of frequency offset between a single frequency laser (SFL) and a mode-locked laser (MLL) less than 350 kHz in 24 hours. Following the incorporation of the analog control component, frequency fluctuation is less than 2.5 Hz in 24 hours. By synchronizing two SFLs to a repetition frequency locked MLL, we achieve indirect synchronization between SFLs with a frequency offset of 10.6 GHz and fluctuation less than 5 Hz in 24-hour, demonstrating robust long-term and short-term stability. Since the MLL is employed as a reference, the system can be utilized for cross-band indirect synchronization of multiple lasers. Based on the synchronization system, we propose a photonic-assisted microwave frequency identification scheme, which has detection error of less than 0.6 MHz. The high performance of the synchronization system enables the proposed frequency identification scheme to achieve high measurement accuracy and theoretically large frequency range.

\end{abstract}


\section{Introduction}
\IEEEPARstart{L}{aser} synchronization system is capable of ensuring that the wavelength of two or more lasers changes in unison, thereby facilitating the stability of the frequency offset between lasers. This technology is widely used in optical communication \cite{harrison1989linewidth,kazovsky19901320}, laser spectroscopy \cite{kaliteevskii1986w}, laser cooling and trapping \cite{metcalf1994cooling}, millimeter- and submillimeter- wave photonics \cite{hyodo2003optical,davidson1998low,simonis1990optical}, hybrid continuous-wave THz imaging \cite{friederich2010phase}, continuous-variable quantum key distribution \cite{laudenbach2019pilot,huang2015high,wang2020high}, injection locking \cite{zitong2018stabilized}. An optical frequency comb (OFC) has an optical spectrum consisting of equidistant lines with a fixed pulse repetition frequency ($f_{rep}$). It can be used as a reference for the single frequency laser (SFL) to achieve frequency stability. When multiple SFLs are synchronized to the OFC at the same time, the frequency offset stabilization of multiple SFLs can be achieved. Due to the broadband optical spectrum characteristic of OFC, the SFLs’ frequency offset can be large enough to be suitable for different applications.
\par The frequency or the phase difference between the SFL and the adjacent reference OFC mode can be controlled via operation temperature, current, and/or the piezo cavity length regulator of the SFL. With the tuning methods described above, the laser synchronization of SFL and OFC can be achieved by digital or analog locking system. However, the digital-only or analog-only locking system has its own shortcomings. The analog locking system is usually based on a analog phase-lock loop (PLL), which has a trade-off between bandwidth and accuracy \cite{citta1977frequency}. And, analog PLL can hardly withstand large frequency abrupt changes \cite{8494198}. Therefore, it is difficult to achieve long-term locking when the laser relative frequency changes rapidly \cite{padgett1988simple,hisai2018evaluation}. The digital locking systems have a relatively large bandwidth, but the speed and accuracy of the feedback device often limit system's locking performance \cite{10056153,jost2002continuously}. The current hybrid "digital + analog" locking systems can combine the advantages of digital and analog locking \cite{couturier2018laser,5446451}. However, in their digital locking section, digital phase frequency detector (PFD) are often used, which have a limited frequency difference between the input signal and the reference signal \cite{mehta2014hybrid,chen2021analysis,7182019}. This may lead to insufficient robustness of the system. Some use frequency counting to discriminate the frequency, the accuracy of this method is not high and the delay is \cite{kratyuk2007frequency,s20051248}. Both of these methods have restrictions on the input signal frequency. 
\par In the above systems, dynamic properties of the control loop should be carefully considered, especially for the response of the laser oscillator and servo actuators \cite{yang2020ultrastable}. However, many commercial SFL cannot be precisely controlled with a bandwidth over the kHz level, which limits the final locking performance. Another promising solution known as the acoustic optical modulator (AOM)-based feedforward method for laser frequency control was applied to synchronization of SFL and OFC \cite{zhang2016linewidth,gatti2012analysis,sala2012wide}. Due to the limited frequency shift range of AOM, the long-term operation of the system becomes difficult. The use of dual-loop "digital+analog" feedback can effectively increase the long-term stability of the system \cite{yang2020ultrastable}. In addition, these systems all introduce another reference source, which will introduce non-common mode noise and thus deteriorate the locking performance. Synchronizing the two reference sources can reduce the non-common mode noise introduced, but it will increase the system complexity.

\par This paper presents a hybrid "digital+analog" laser synchronization system for synchronizing SFLs with a $f_{rep}$ locking MLL. The analog locking part extracts the $f_{rep}$ harmonics, and through the signal mixing and filtering, the error signal is directly loaded on the acousto-optic modulator (AOM) to reduce the non-common mode noise. In digital locking part, we propose the electric intensity method, which utilizes a combination of a bandpass filter (BPF) and an RMS detector for frequency discrimination. The method has low delay and large bandwidth, ensures the speed of feedback and improves the long-term stability of locking. In addition, there is no need to introduce microwave sources in the digital or analog locking parts, which reduces the complexity of the system. With digital locking only, the fluctuation of frequency offset between an MLL and an SFL is less than 350 kHz in 24 hours. Following the activation of the analog locking, the fluctuation was less than 2.5 Hz. We indirectly synchronized two SFLs with a frequency offset of 0.9 GHz, 4.9 GHz and 10.6 GHz, and a 24-hour frequency fluctuation of 4 Hz, 4.7 Hz, and 5 Hz respectively. Based on the synchronization system, we propose a high measurement accuracy comb-assisted frequency identification scheme, which has detection error of less than 0.6 MHz. Due to the wide spectrum characteristic of MLL, the system is theoretically capable of measuring a wide range of frequencies.

\section{Hybrid "digital + analog" single-branch laser synchronization system}
\subsection{System Design}

\begin{figure*}[ht!]
\centering
\includegraphics[width=0.8\textwidth]{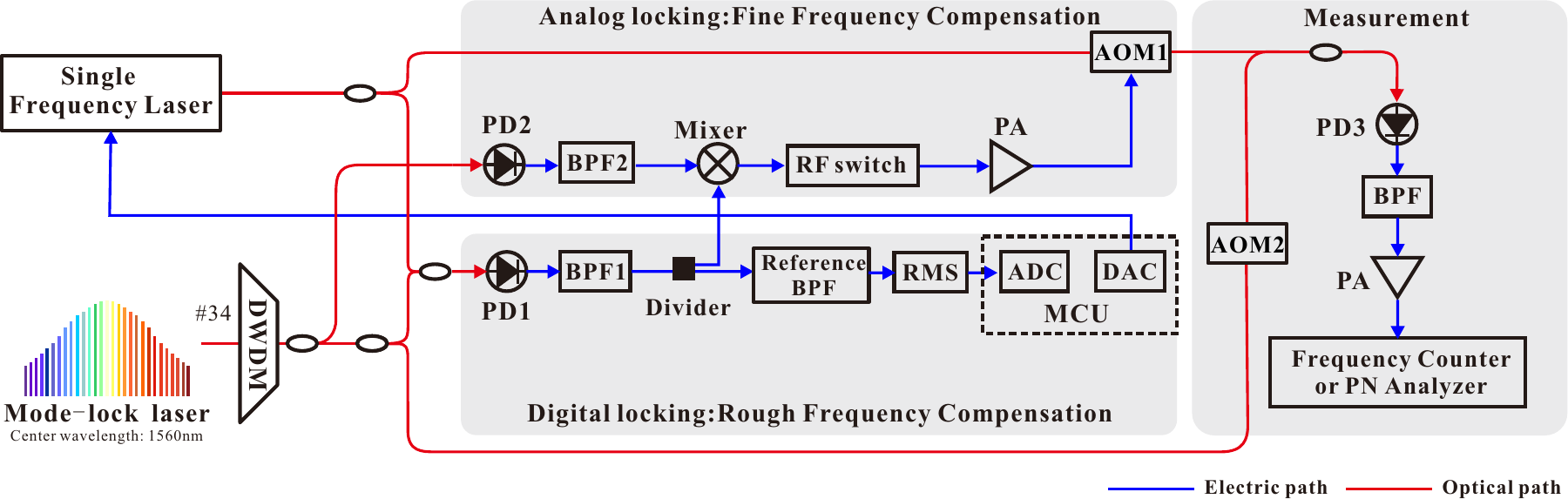}
\caption{The schematic of single-branch laser synchronization system. DWDM: dense wavelength division multiplexing, BPF: band-pass filter, PD: photodetector, Divider: power divider, RF Switch: radio frequency switch, AOM: acousto-optic modulator, RMS: root mean square detector, ADC: analog-to-digital converter, DAC: digital-to-analog converter, MCU: microcontroller unit, PA: power amplifier, PN Analyzer: Phase noise analyzer.}
\end{figure*}
Fig.1(a) shows the schematic of the single-branch laser synchronization system. The $f_{rep}$ of the MLL (homemade) is 100 MHz, and the central wavelength is 1560.2 nm. The $f_{rep}$ of the MLL is locked but $f_{ceo}$ is unlocked. The center wavelength of the single frequency laser (SFL) is 1550.12 nm. In the digital locking, the output light of MLL is filtered by the 34 channels of the DWDM and coupled to the SFL through a 50:50 coupler, and then enters photodetector1 (PD1). The beat frequency ($f_{beat}$) signal is obtained after BPF1. After passing through the power divider, one branch signal travels to the analog locking and another enter into reference BPF. The RMS detector, following reference BPF, detects signal intensity and converts it to voltage. The ADC is used to get the output voltage of the RMS detector and the MCU is used to read the voltage and calculate the feedback voltage, which is fed back to the piezoelectric transducer (PZT) of the SFL through the DAC. In the analog locking, the third harmonic of $f_{rep}$ is extracted by PD2 and BPF2, then mixed with the $f_{beat}$ of SFL and MLL, which is from the power divider in digital locking. After the mixer, the signal enters RF switch to select up-conversion or down-conversion signals to pass. Subsequently, the selected signal is loaded on the AOM1 in order to achieve frequency compensation following the electrical amplifier. Since the SFL and MLL are homodyne synchronized, the locking error is difficult to measure directly. Therefore, we add AOM2 in the measurement system to generate a fixed frequency shift. The frequency counter is Keisight 53230A. The phase noise (PN) analyzer is Rohde $\&$ Schwarz FSPN.

\subsection{Digital locking based on electrical intensity}
\begin{figure}[t]
\centering
\includegraphics[width=0.3\textwidth]{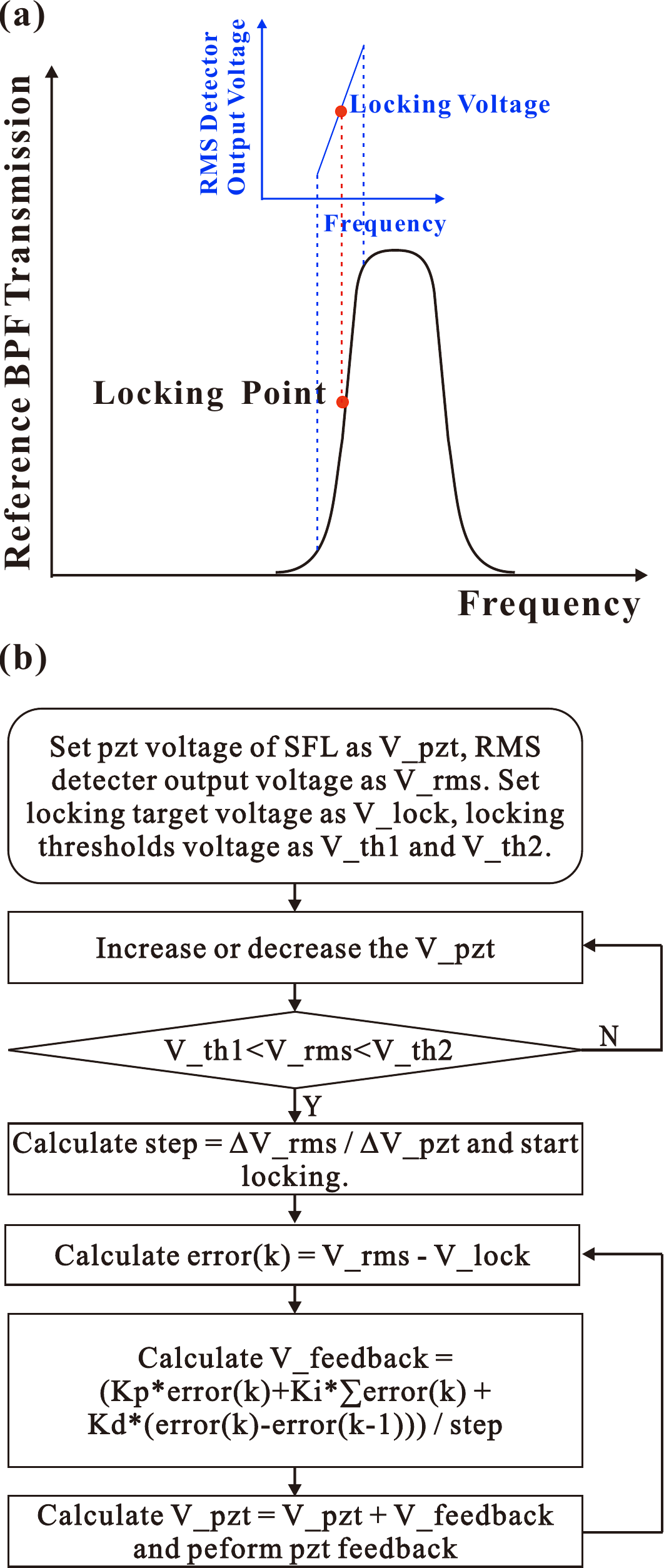}
\caption{(a) Principle of electrical intensity locking in digital locking: The combination of BPF and RMS detector realizes that frequency to voltage. Based on this voltage, the laser frequency offset is locked on a roll-off band of BPF. (b) Flowchart of the digital locking algorithm.}
\end{figure}
We present the electrical intensity locking technique based on BPF and RMS detector. The schematic of the technique is shown in Fig. 2(a). The solid black line represents the filter transmission versus signal frequency, and the solid blue line represents the root mean square (RMS) detector output voltage versus signal frequency. When the frequency of the input signal is at the highest point of filter transmittance, the RMS detector output voltage is maximum, and when the filter is not at the highest point of transmittance, the RMS detector output voltage will decrease accordingly. When the input signal frequency is in the filter roll-off band, the signal frequency and RMS detector output voltage are approximately linear, so we can use this technique to achieve frequency locking. We use this technique in the digital locking locking in Fig. 1(a). The reference BPF is a narrow-band filter with a center frequency of 35 MHz. In order to improve the sensitivity of the feedback and thus improve the locking accuracy, we choose the maximum slope point as the locking point, the point corresponds to the RMS detector output voltage is the locking target voltage. Fig. 2(b) shows the flow chart of electrical intensity frequency locking algorithm. Firstly, the laser wavelength is made to move near the locking point, and then the error with the target locking voltage is calculated based on the current RMS detector output voltage, and the feedback voltage is calculated by the PID algorithm. The feedback voltage is loaded to the PZT of the SFL by the MUC control ADC, and the PID parameters can be adjusted to get a good locking effect.
\par Fig. 3(a) shows the variation in frequency offset of SFL and MLL over 12 hours without synchronization. Since the $f_{rep}$ of of the MLL is 100 MHz, the maximum range of frequency offset variation is theoretically $0\sim50$ MHz. Due to limited measurement accuracy, the offset frequency we detect varies between 5 MHz and 45 MHz. After only digital locking, the variation of frequency offset of MLL and SFL within 350 kHz and its RSME is 51.81 kHz over 24 hours in Fig. 3(b). The results show that the locking based on the electrical strength method has good long-term stability. 
\begin{figure}[t!]
\centering
\includegraphics[width=0.47\textwidth]{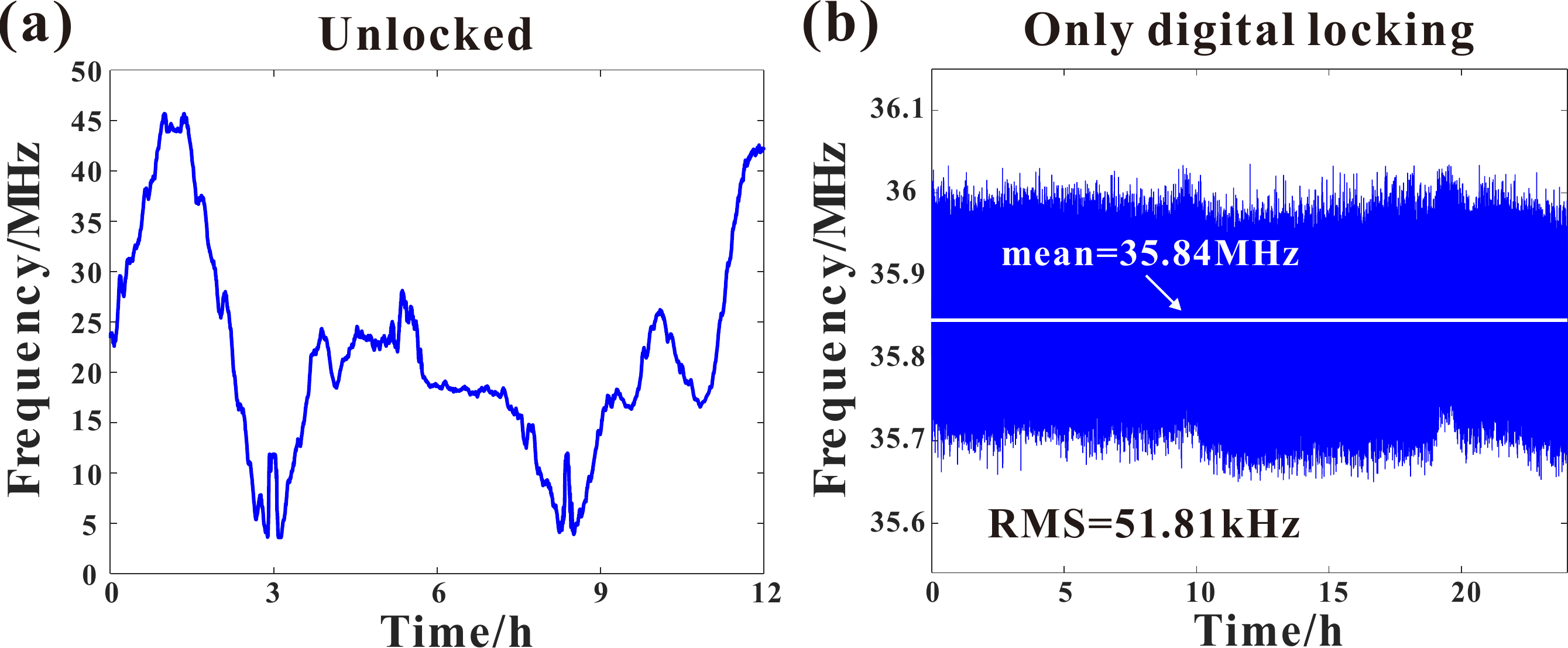}
\caption{(a) Frequency offset between SFL and MLL in a free-running state for 12 hours. (b) Laser frequency offset in 24 hours with only digital locking.}
\end{figure}
\subsection{Analog locking based on mixing and filter}
\begin{figure}[t!]
\centering
\includegraphics[width=0.47\textwidth]{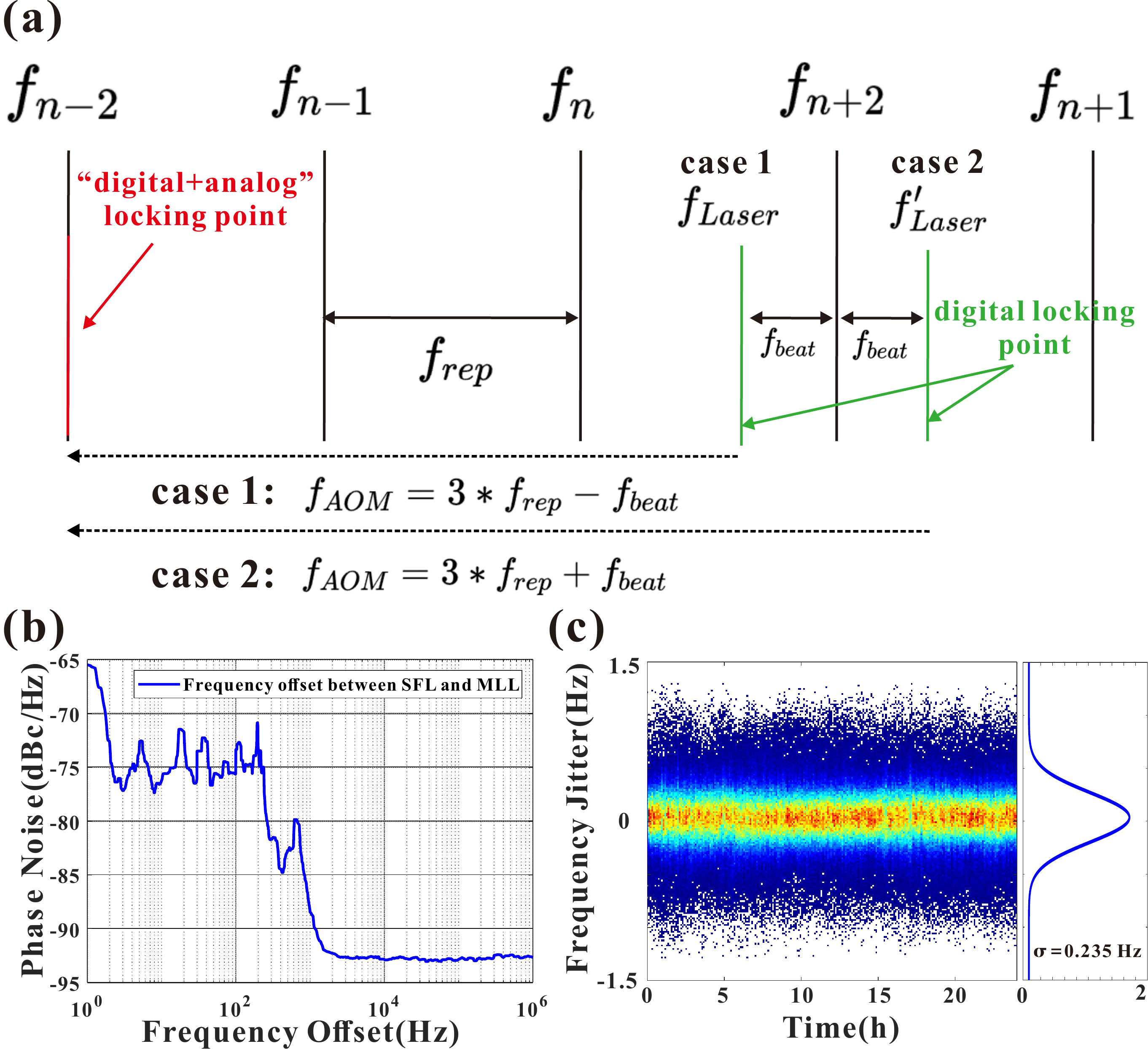}
\caption{(a) Principle of the homodyne laser synchronization with "digital+analog" locking. Black lines: optical modes of MLL, Green lines :two cases of SFL frequency after only digital locking, Red line: SFL frequency after hybrid "digital+analog" synchronization. (b) Phase noise of frequency offset after hybrid synchronization (c) The density distribution of relative frequency fluctuation bwteen SFL and MLL in 24 hours after hybrid synchronization. $\sigma$ = 0.235 Hz.}
\end{figure}
In Fig.1(a)'s analog locking part, PD2 and BPF2 extract the third harmonic of $f_{rep}$, which mixes with the $f_{beat}$ signal from digital locking part. The reason for using third harmonic of $f_{rep}$ mixing is that we are using the AOM1 control range of 300 MHz $\pm$ 50 Mhz. The mixer output divides into two paths, each passing through a filter with a distinct center frequency. The appropriate path is selected by the radio frequency (RF) switch. Fig.4(a) illustrates the principle of RF path selection. The solid black lines represent the optical modes of the MLL. The solid green lines $f_{Laser}$ and $f_{Laser}^{'}$ represent two cases of SFL frequency after only digital locking. The solid red line represents SFL frequency after "digital+analog" locking, which coincides exactly with another optical mode of the MLL. In case 1, the $f_{Laser}$ is less than $f_{n+2}$ and therefore the RF switch selection is $f_{AOM}=3*f_{rep}-f_{beat}$. In case 2, the $f_{Laser}^{'}$ is more than $f_{n+2}$, resulting in the selection of $f_{AOM}=3*f_{rep}+f_{beat}$. The mixed signal is then amplified, and loaded onto AOM1 to achieve homodyne frequency compensation. The phase noise of the SFL and MLL frequency difference after locking is shown in Fig. 4(b). The phase noise exhibits up-and-down fluctuations, which we believe is caused by the delay compensated by the synchronization system. Despite the $f_{rep}$ of MLL locked, there is still a small fluctuation, which is amplified by approximately 1.92 million times when it is transferred from the first mode of MLL to the mode to be synchronized. The delay of the optical fiber and electrical components in the synchronization system results in a lag in the feedback, which in turn leads to the fluctuation not being fully compensated. Fig. 4(c) illustrates relative frequency fluctuation between SFL and MLL after hybrid "digital + analog" synchronization in 24 hours. The RMS error is 0.23 Hz. Table 1 summarizes the reported synchronization techniques for SFL with MLL, indicating that our hybrid synchronization system provides the short-term stability of analog locking with the long-term stability of digital locking. 
\begin{table*}
\centering
\caption{Comparison of the Single Frequency Laser Synchronized to Mode-Locked Laser or Optical Frequency Comb}
\begin{tabular}{cccc}
\hline % 一条横线
Works & Frequency fluctuation & Beat Frequency Stability (Short-Term) & Beat Frequency Stability (Long-Trem)\\

\hline % 一条横线
Ref. \cite{s20051248} & Below 13.94Hz@10h& 8.01E-16@10s & 2.19E-16@1000s \\
\hline % 一条横线
Ref. \cite{fang2015coherence} & - & 1.3E-15@1s & - \\
\hline % 一条横线
Ref. \cite{mcferran2018laser}& - & 1.3E-13@1s & - \\
\hline % 一条横线mcferran2018laser
Ref. \cite{zhang2022frequency}& Below 2Hz@4000s & 3.5E-16@1s & - \\
\hline % 一条横线
Ref. \cite{argence2015quantum}& Below 300Hz@3500s & 2E-16@1s & 1E-17@100s \\
\hline % 一条横线
Ref. \cite{yasui2021precise}& Below 300Hz@3500s & - & - \\
\hline % 一条横线
Ref. \cite{miyashita2021offset}& Below 30kHz@3000s & 3.6E-12@1s & - \\
\hline % 一条横线
Ref. \cite{Zhou:21} & Below 1MHz@4000s & - & 3.5E-12@4000s \\
\hline % 一条横线
Ref. \cite{9511311} & Below 3Hz@100000s & - & 3.6E-18@10000s \\
\hline % 一条横线

This work & Below 2.5Hz@24h & 1.71E-16@1s & 2.101E-20@10000s \\
\hline % 一条横线
\end{tabular}
\end{table*}

\section{Hybrid "digital + analog" multi-branch laser synchronization system}
\subsection{System Design}
\begin{figure*}[ht!]
\centering
\includegraphics[width=1\textwidth]{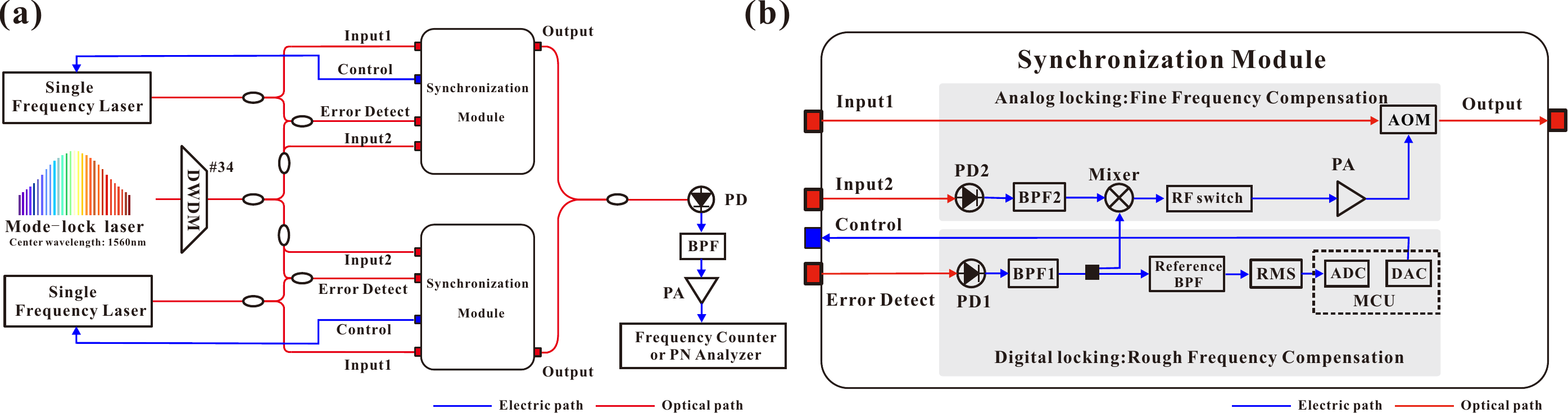}
\caption{(a) The schematic of multi-branch laser synchronization system. (b) The schematic of synchronization module.}
\end{figure*}
By copying multiple single-branch synchronization systems, we can achieve multiple SFLs frequency synchronization. Here, we take the example of a two-branch SFL synchronization. Fig. 5(a) shows the scheme of a two-branch frequency synchronization system. The synchronization module details are presented in Fig. 5(b), and the locking principle is identical to that of the single-branch synchronization system. 
\subsection{Experimental Results}
\begin{figure}[ht!]
\centering
\includegraphics[width=0.48\textwidth]{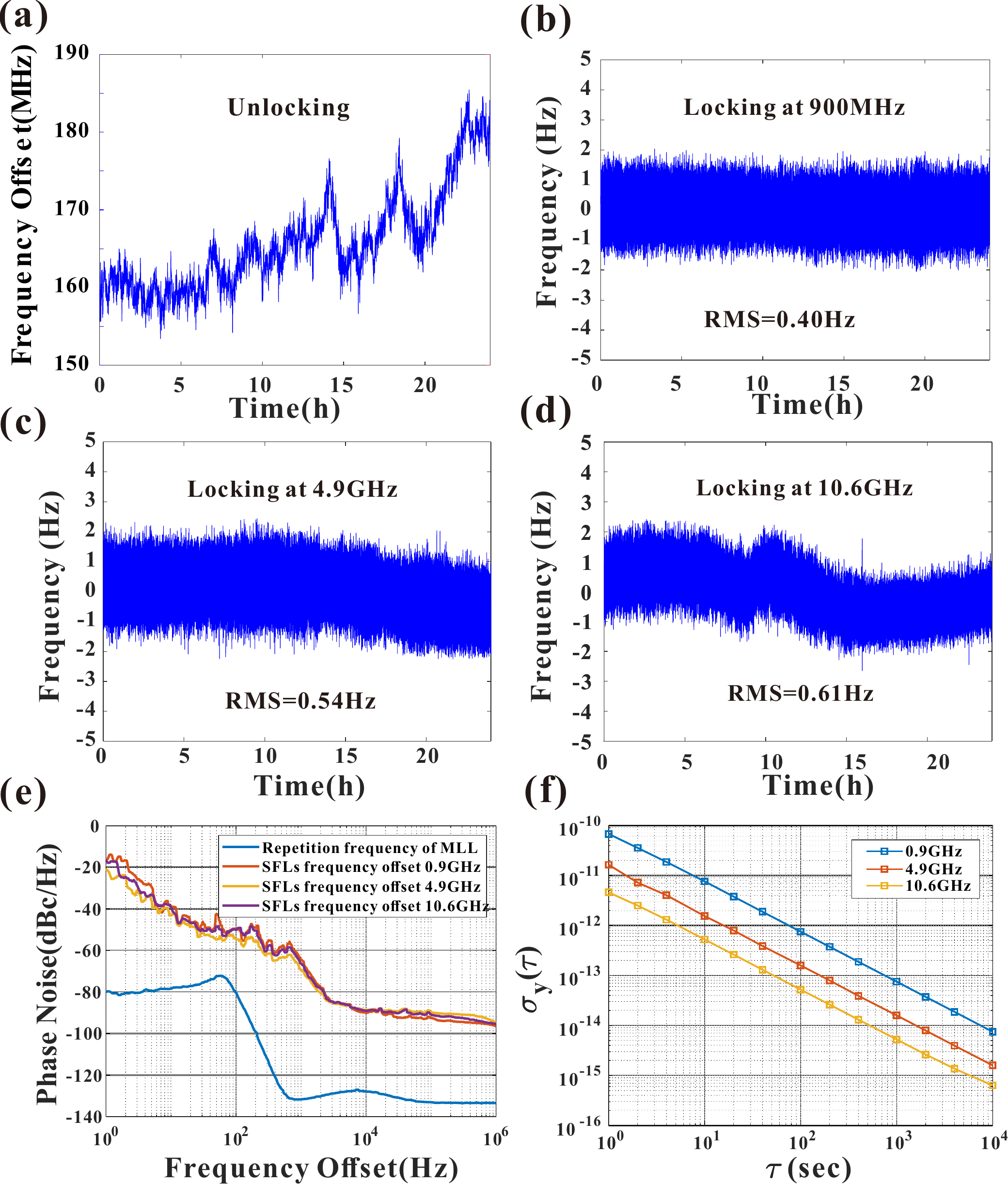}
\caption{(a) Frequency offset of SFLs when unlocked. (b) Relative frequency fluctuation of SFLs locking at 0.9 GHz, (c)locking at 4.9 GHz, (d)locking at 10.6 GHz. (e)The phase noise of SFLs' frequency offsets and MLL's $f_{rep}$ after synchronization. (f)The ADEV of SFLs' laser frequency offsets after synchronization.}
\end{figure}
Fig. 6(a) shows the frequency offset variation of two SFLs when SFLs and MLL are not synchronized. The variation is approximately 30 MHz over a 24-hour period. We adjusted the initial frequency offsets of the two SLFs to 0.9 GHz, 4.9 GHz, and 10.6 GHz respectively then synchronized the SFLs to an MLL, and the detail results obtained are shown in Table 2. In Fig. 5(c) and (d), the relative frequency fluctuation at 4.9 GHz and 10.6 GHz have a long-term trend, which may be due to the environmental influence of the homemade $f_{rep}$ locking system, resulting in minor changes in the MLL $f_{rep}$. Fig. 6(e) illustrates the phase noise of MLL's $f_{rep}$ and SFLs' different frequency offsets after synchronization. Compared to MLL's $f_{rep}$, the phase noise of the SFLs' frequency offset after synchronization is much worse. Taking the 4.9 GHz offset frequency as an example, the offset frequency are equivalent to 49 times the $f_{rep}$ when synchronizing, and two synchronization systems are introduced at the same time, leading to a deterioration of the stability. Furthermore, when detecting the SLFs' beat frequency, due to the low RF power directly detected by PD, the phase noise analyzer is unable to measure, necessitating the addition of power amplifier, which will inevitably introduce noise. Fig. 6(f) demonstrates that the ADEV at different offset frequencies all exhibit a drop in $\frac{1}{\tau } $. The advantage of this system is that the system structure is simple and it can lock multiple SFLs wavelengths in any MLL wavelength coverage range, and if the MLL is spectrally broadened, the locking of the laser frequency offset for ultra-far spanning bands can be realized.
\begin{table}[!htbp]
\centering
\caption{Results of 24-hour synchronization at different laser offset frequency.}
\begin{tabular}{cccc}
\hline % 一条横线
Laser offset frequency (GHz) & 0.9 & 4.9 & 10.6 \\
\hline % 一条横线
Peak to peak error (Hz) & 4 & 4.7 & 5 \\
\hline % 一条横线
RMSE (Hz) & 0.40 & 0.54 & 0.61 \\
\hline % 一条横线
ADEV@1s & 6.79E-11 & 1.64E-11 & 4.61E-12 \\
\hline % 一条横线
ADEV@10000s & 7.49E-15 & 1.59E-15 & 6.34E-16 \\
\hline % 一条横线
\end{tabular}
\end{table}
\section{Comb-assisted microwave frequency identification}
\subsection{System Design and Principle}

\begin{figure*}[ht!]
  \centering
  \includegraphics[width=0.8\textwidth]{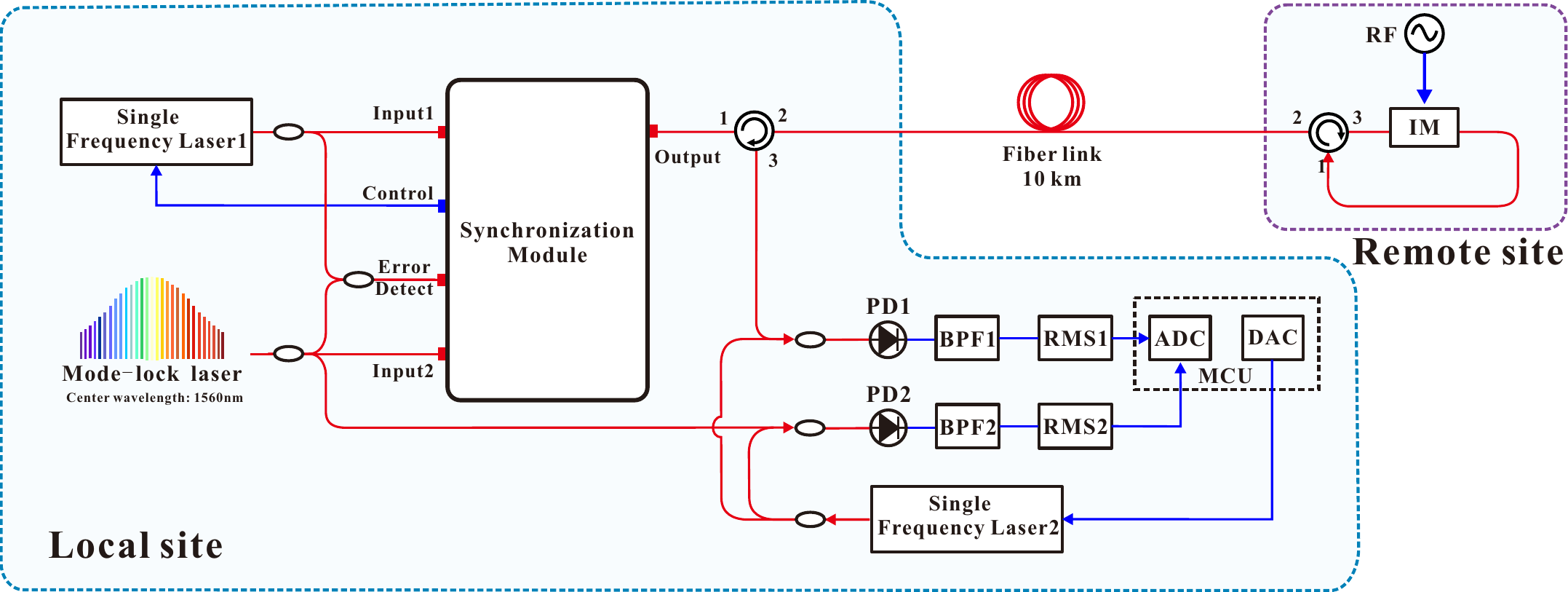}
\caption{Schematic diagram of the experimental setup. BPF: band pass filter, RMS: root mean square detector, MCU: micro-controller unit, IM: Intensity modulator}
\end{figure*}
\begin{figure*}[h!]
  \centering
  \includegraphics[width=1\textwidth]{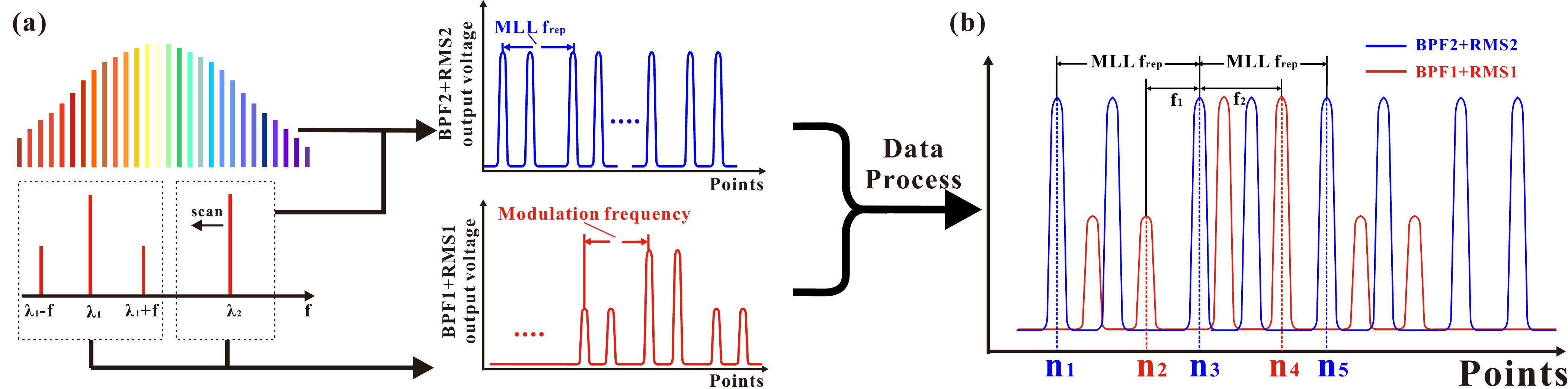}
\caption{The principle of comb-assisted microwave frequency identification.}
\end{figure*}
As shown in Fig.7, SFL1 is synchronized on a $f_{rep}$ stabilized MLL by synchronization module (same as synchronization module in Fig.5(b)). The MLL is the same MLL as the one in Fig. 5, and the $f_{rep}$ is 100 MHz. The synchronized light is transmitted to the remote site through a 10 km optical fiber. To simulate the unknown microwave signal received by the antenna, and the RF signal of 445 MHz and 925 MHz is modulated by the intensity modulator (IM) in the remote site, which is the signal to be measured. The signal at the remote end is transmitted back and coupled with the output light of SFL2 at the local end to enter PD1. Meanwhile, the MLL on the local side is coupled with the output of SFL2 into PD2. A DAC is used to control the wavelength scanning of SFL2, and an ADC with dual input ports is used to record the output voltage values of RMS detector 1 and RMS detector 2 simultaneously. BPF1 and BPF2 are bandpass filters with a central wavelength of 35 MHz and a 3 dB bandwidth of 4 MHz. 
\par The measurement principle is shown in Fig. 8(a). After SLF1 ($\lambda _{1} $) is synchronized with MLL, the wavelength of SLF2 ($\lambda _{2} $) starts being scanned. The output of SLF2 is coupled with MLL and modulated SFL1, respectively. After photoelectric detection ,the signals goes into different BPFs and RMS detectors. The output voltage of the RMS detectors is recorded simultaneously. The blue line represents the output voltage of RMS detector 2, the red line represents the output voltage of RMS detector 1, and the abscissa is both the number of points acquired by the ADC. The RMS detectors output has many voltage peaks, which are caused by the beat signal passing through the BPF. The voltage peaks appear in pairs, which is due to the fact that a voltage peak is generated on both the left and right sides of $\lambda _{2} $ passing through a certain optical frequency during frequency scaning. The distance between a pair of peaks represents twice the center frequency of the BPF. In the output voltage of RMS detector 2, the distance between each pair of peaks represents the $f_{rep}$ of the MLL. In the output voltage of RMS detector 1, the pair of peaks with the highest voltage represents the carrier, the pair of peaks with the low voltage represents the modulated sidebands, and the distance between the two pairs of peaks represents the frequency of the modulated signal to be measured. The RMS detector 1 and RMS detector 2 output voltages are acquired simultaneously by a dual-input ADC, so their data strictly correspond in the time dimension, shown in Fig. 8(b). The $f_{rep}$ of MLL can be obtained by frequency counter measurements. According to the position of the peak point of RMS detector 1 and RMS detector 2 and the ratio of the number of points between them, the value of $f_{1}$ and $f_{2}$ can be obtained, and the modulation frequency can be calculated as $f_{1}+f_{2}$. As shown in Fig. 8 (b), $n_{1}\sim n_{5}$ represent the number of data points collected when the RMS detector voltage is at different peak points. Then $f_{1}$ can be expressed as
\begin{equation}
\label{deqn_ex1a}
f_{1} = f_{rep}\frac{n_{3} -n_{2} }{n_{3} -n_{1}} 
\end{equation}
, and $f_{2}$ can be expressed as
\begin{equation}
\label{deqn_ex1a}
f_{2} = f_{rep}\frac{n_{4} -n_{3} }{n_{5} -n_{3}}  
\end{equation}
Fig. 8 (b) shows the case when the modulation frequency is less than the MLL's $f_{rep}$. When the modulation frequency is greater than the $f_{rep}$, it can be expressed as $f_{1}+f_{2}+nf_{rep}$, where n is a positive integer.
\subsection{Experimental Results}
\begin{figure}[t!]
  \centering
   \includegraphics[width=0.47\textwidth]{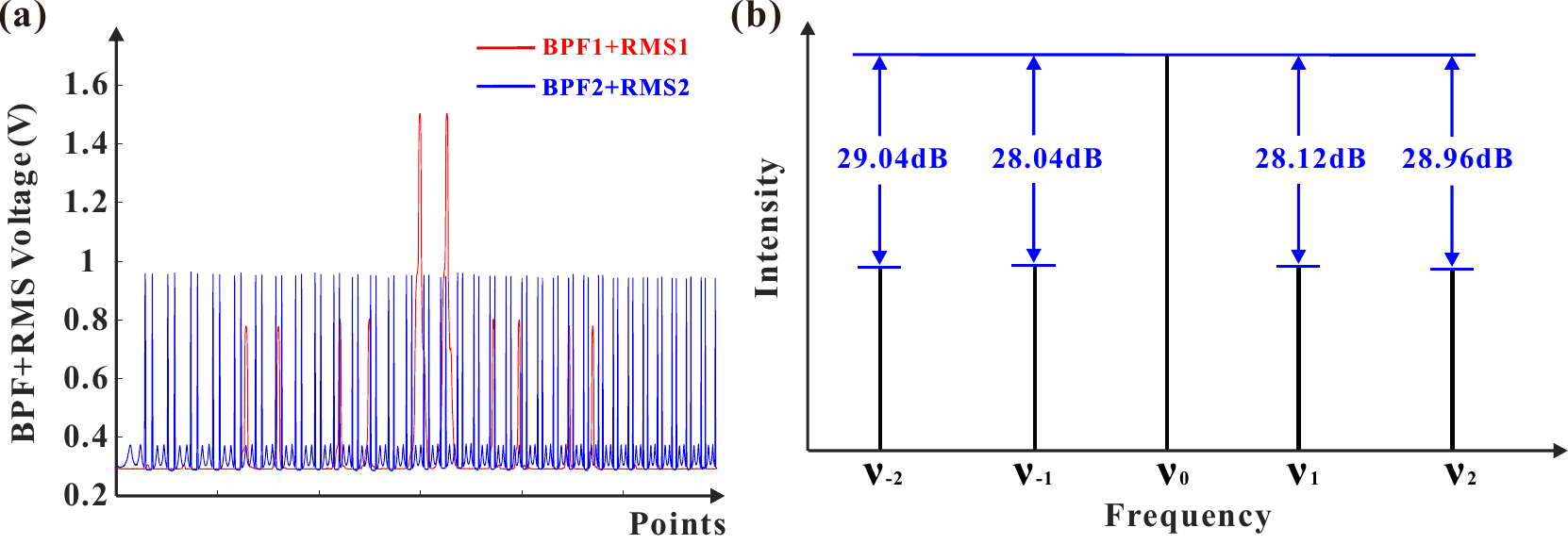}
    \caption{ (a) The actual output voltages of RMS detector 1 and RMS detector 2. (b) Recovered optical spectrum.}
\end{figure}

\begin{table*}[b]
\centering
\caption{Measurement errors at different modulation frequencies}
\begin{tabular}{ccccccc}
\hline % 一条横线
Modulation frequency (MHz) & 300  & 384 & 445 & 713 & 807  & 1000 \\
\hline % 一条横线
Measurement result (MHz) & 300.19  & 383.76 & 445.11 & 712.45 & 807.20  & 999.50 \\
\hline % 一条横线
Error (MHz) & 0.19  & 0.24 & 0.11 & 0.55 & 0.20  & 0.50 \\
\hline % 一条横线
\end{tabular}
\end{table*}

Fig. 9(a) shows the RMS detectors voltage of the actual acquisition. The optical power ratio between the carrier and the modulated sideband can be calculated as follows:
\begin{equation}
\label{deqn_ex1a}
Ratio = (V_{carrier}-V_{modulation})Slope.
\end{equation}
where $V_{carrier}$ and $V_{carrier}$ are the voltage of the carrier peak and the voltage of the modulated sideband peak in the RMS detector 1 output, respectively. $Slope$ is the logarithmic slope of the RMS detector in mV/dBm. Fig. 9(b) illustrates the recovered optical spectral sketch, where $\nu_{0} $ represents the carrier. $\nu_{-1} $ and $\nu_{1} $ represent the first pair of modulation sidebands, which are calculated as $\nu_{0}$ - 445.16 MHz and $\nu_{0}$ + 445.06 MHz respectively. After averaging, the modulation frequency is 445.11 MHz, and the error is 0.11 Mhz with the actual modulation frequency. $\nu_{-2} $ and $\nu_{2} $ represent the second pair of modulation sidebands, which are calculated as $\nu_{0}$ - 924.76 MHz, $\nu_{0}$ - 925.24 MHz, respectively. After averaging, the modulation frequency is 925 MHz, and the error with the actual modulation frequency is 0 Mhz. We measured the modulated signal at multiple frequencies as shown in Table 3, with the highest frequency being 1GHz due to the limitation of the bandwidth of the IM. In Table 3, the measurement error is up to 0.55 MHz and does not increase as the modulation frequency increases. The frequency identification error of the system mainly includes the synchronization error between SFL1 and MLL and the accuracy and speed of ADC and DAC. Due to the introduction of the synchronization system, the SFL1 and MLL frequency fluctuation is less than 2.5 Hz. The accuracy and speed of the ADC and DAC will influence the determination of the voltage peak point, potentially leading to the generation of errors. Due to the wide spectral characteristics of MLL, the theoretical measurement range of the system is the wavelength tunable range of SFL. 
\section{Conclusion}
This paper presents a hybrid "digital + analog" laser synchronization system with low-complexity and high-performance, realizing the fluctuation of frequency offset between a single frequency laser (SFL) and a mode-locked laser (MLL) less than 2.5 Hz in 24 hours. The proposed scheme avoids the use of an RF reference source, simplifying the loop structure while providing enhanced short- and long-term stability. By synchronizing two SFLs to an MLL, we achieve indirect synchronization between the SFLs. Since the MLL is employed as a reference, the system can be utilized for cross-band indirect synchronization of multiple lasers. If the MLL is replaced by the fully locked OFC, the frequency stabilization of multiple SFLs can be realized. Based on the synchronization system, we propose a photonic-assisted microwave frequency identification scheme, which has detection error of less than 0.6 MHz. The high performance of the synchronization system enables the proposed frequency identification scheme to achieve high measurement accuracy and a theoretically large frequency range.
\bibliographystyle{IEEEtran}
\bibliography{sample}

\begin{thebibliography}{10}
\providecommand{\url}[1]{#1}
\csname url@samestyle\endcsname
\providecommand{\newblock}{\relax}
\providecommand{\bibinfo}[2]{#2}
\providecommand{\BIBentrySTDinterwordspacing}{\spaceskip=0pt\relax}
\providecommand{\BIBentryALTinterwordstretchfactor}{4}
\providecommand{\BIBentryALTinterwordspacing}{\spaceskip=\fontdimen2\font plus
\BIBentryALTinterwordstretchfactor\fontdimen3\font minus \fontdimen4\font\relax}
\providecommand{\BIBforeignlanguage}[2]{{%
\expandafter\ifx\csname l@#1\endcsname\relax
\typeout{** WARNING: IEEEtran.bst: No hyphenation pattern has been}%
\typeout{** loaded for the language `#1'. Using the pattern for}%
\typeout{** the default language instead.}%
\else
\language=\csname l@#1\endcsname
\fi
#2}}
\providecommand{\BIBdecl}{\relax}
\BIBdecl

\bibitem{harrison1989linewidth}
J.~Harrison and A.~Mooradian, ``Linewidth and offset frequency locking of external cavity gaalas lasers,'' \emph{IEEE journal of quantum electronics}, vol.~25, no.~6, pp. 1152--1155, 1989.

\bibitem{kazovsky19901320}
L.~G. Kazovsky and D.~A. Atlas, ``A 1320-nm experimental optical phase-locked loop: performance investigation and psk homodyne experiments at 140 mb/s and 2 gb/s,'' \emph{Journal of Lightwave Technology}, vol.~8, no.~9, pp. 1414--1425, 1990.

\bibitem{kaliteevskii1986w}
N.~Kaliteevskii, ``W. demtr{\"o}der, laser spectroscopy: Basis concepts and instrumentation (springer-verlag, heidelberg, 1982; nauka, moscow 1985),'' \emph{Optics and Spectroscopy}, vol.~61, no.~1, p. 134, 1986.

\bibitem{metcalf1994cooling}
H.~Metcalf and P.~van~der Straten, ``Cooling and trapping of neutral atoms,'' \emph{Physics reports}, vol. 244, no. 4-5, pp. 203--286, 1994.

\bibitem{hyodo2003optical}
M.~Hyodo and M.~Watanabe, ``Optical generation of millimeter-wave signals up to 330 ghz by means of cascadingly phase locking three semiconductor lasers,'' \emph{IEEE Photonics Technology Letters}, vol.~15, no.~3, pp. 458--460, 2003.

\bibitem{davidson1998low}
A.~C. Davidson, F.~W. Wise, and R.~C. Compton, ``Low phase noise 33-40-ghz signal generation using multilaser phase-locked loops,'' \emph{IEEE Photonics Technology Letters}, vol.~10, no.~9, pp. 1304--1306, 1998.

\bibitem{simonis1990optical}
G.~J. Simonis and K.~G. Purchase, ``Optical generation, distribution, and control of microwaves using laser heterodyne,'' \emph{IEEE Transactions on microwave theory and techniques}, vol.~38, no.~5, pp. 667--669, 1990.

\bibitem{friederich2010phase}
F.~Friederich, G.~Schuricht, A.~Deninger, F.~Lison, G.~Spickermann, P.~H. Bol{\'\i}var, and H.~G. Roskos, ``Phase-locking of the beat signal of two distributed-feedback diode lasers to oscillators working in the mhz to thz range,'' \emph{Optics express}, vol.~18, no.~8, pp. 8621--8629, 2010.

\bibitem{laudenbach2019pilot}
F.~Laudenbach, B.~Schrenk, C.~Pacher, M.~Hentschel, C.-H.~F. Fung, F.~Karinou, A.~Poppe, M.~Peev, and H.~H{\"u}bel, ``Pilot-assisted intradyne reception for high-speed continuous-variable quantum key distribution with true local oscillator,'' \emph{Quantum}, vol.~3, p. 193, 2019.

\bibitem{huang2015high}
D.~Huang, P.~Huang, D.~Lin, C.~Wang, and G.~Zeng, ``High-speed continuous-variable quantum key distribution without sending a local oscillator,'' \emph{Optics letters}, vol.~40, no.~16, pp. 3695--3698, 2015.

\bibitem{wang2020high}
H.~Wang, Y.~Pi, W.~Huang, Y.~Li, Y.~Shao, J.~Yang, J.~Liu, C.~Zhang, Y.~Zhang, and B.~Xu, ``High-speed gaussian-modulated continuous-variable quantum key distribution with a local local oscillator based on pilot-tone-assisted phase compensation,'' \emph{Optics express}, vol.~28, no.~22, pp. 32\,882--32\,893, 2020.

\bibitem{zitong2018stabilized}
F.~Zitong, Y.~Fei, Z.~Xi, C.~Dijun, C.~Nan, G.~YouZhen, and C.~Haiwen, ``Stabilized optical-frequency transfer using optical injection locking amplifier,'' in \emph{2018 European Frequency and Time Forum (EFTF)}.\hskip 1em plus 0.5em minus 0.4em\relax IEEE, 2018, pp. 178--180.

\bibitem{citta1977frequency}
R.~Citta, ``Frequency and phase lock loop,'' \emph{IEEE Transactions on Consumer Electronics}, no.~3, pp. 358--365, 1977.

\bibitem{8494198}
A.~Godave, P.~Choudhari, and A.~Jadhav, ``Comparison and simulation of analog and digital phase locked loop,'' in \emph{2018 9th International Conference on Computing, Communication and Networking Technologies (ICCCNT)}, 2018, pp. 1--4.

\bibitem{padgett1988simple}
M.~Padgett, N.~Bett, and R.~Butcher, ``A simple frequency discriminator circuit for offset locking of lasers,'' \emph{Journal of Physics E: Scientific Instruments}, vol.~21, no.~6, p. 554, 1988.

\bibitem{hisai2018evaluation}
Y.~Hisai, K.~Ikeda, H.~Sakagami, T.~Horikiri, T.~Kobayashi, K.~Yoshii, and F.-L. Hong, ``Evaluation of laser frequency offset locking using an electrical delay line,'' \emph{Applied Optics}, vol.~57, no.~20, pp. 5628--5634, 2018.

\bibitem{10056153}
S.~Ding, C.~Wu, E.~Zhu, J.~Shang, T.~Jiang, B.~Luo, and S.~Yu, ``Neural network assisted laser frequency locking system,'' \emph{Journal of Lightwave Technology}, pp. 1--7, 2023.

\bibitem{jost2002continuously}
J.~D. Jost, J.~L. Hall, and J.~Ye, ``Continuously tunable, precise, single frequency optical signal generator,'' \emph{Optics express}, vol.~10, no.~12, pp. 515--520, 2002.

\bibitem{couturier2018laser}
L.~Couturier, I.~Nosske, F.~Hu, C.~Tan, C.~Qiao, Y.~Jiang, P.~Chen, and M.~Weidem{\"u}ller, ``Laser frequency stabilization using a commercial wavelength meter,'' \emph{Review of Scientific Instruments}, vol.~89, no.~4, 2018.

\bibitem{5446451}
T.~T.-Y. Lam, B.~J.~J. Slagmolen, J.~H. Chow, I.~C.~M. Littler, D.~E. McClelland, and D.~A. Shaddock, ``Digital laser frequency stabilization using an optical cavity,'' \emph{IEEE Journal of Quantum Electronics}, vol.~46, no.~8, pp. 1178--1183, 2010.

\bibitem{mehta2014hybrid}
M.~Mehta and V.~Chandrasekhar, ``A hybrid analog-digital phase-locked loop for frequency mode non-contact scanning probe microscopy,'' \emph{Review of Scientific Instruments}, vol.~85, no.~1, 2014.

\bibitem{chen2021analysis}
C.~Chen, P.~Ye, S.~Liao, L.~Xu, J.~Zhang, and F.~Tan, ``Analysis and modeling of hybrid analog-digital pll,'' in \emph{2021 IEEE 15th International Conference on Electronic Measurement \& Instruments (ICEMI)}.\hskip 1em plus 0.5em minus 0.4em\relax IEEE, 2021, pp. 332--335.

\bibitem{7182019}
S.~M. Jung and J.~M. Roveda, ``A low jitter digital phase-locked loop with a hybrid analog/digital pi control,'' in \emph{2015 IEEE 13th International New Circuits and Systems Conference (NEWCAS)}, 2015, pp. 1--4.

\bibitem{kratyuk2007frequency}
V.~Kratyuk, P.~Hanumolu, U.~Moon, and K.~Mayaram, ``Frequency detector for fast frequency lock of digital plls,'' \emph{Electronics Letters}, vol.~43, no.~1, p.~1, 2007.

\bibitem{s20051248}
\BIBentryALTinterwordspacing
R.~Yang, H.~Lv, J.~Luo, P.~Hu, H.~Yang, H.~Fu, and J.~Tan, ``Ultrastable offset-locking continuous wave laser to a frequency comb with a compound control method for precision interferometry,'' \emph{Sensors}, vol.~20, no.~5, 2020. [Online]. Available: \url{https://www.mdpi.com/1424-8220/20/5/1248}
\BIBentrySTDinterwordspacing

\bibitem{yang2020ultrastable}
------, ``Ultrastable offset-locking continuous wave laser to a frequency comb with a compound control method for precision interferometry,'' \emph{Sensors}, vol.~20, no.~5, p. 1248, 2020.

\bibitem{zhang2016linewidth}
Z.~Zhang, Y.~Dai, P.~Ou, F.~Yin, Y.~Zhou, J.~Li, and K.~Xu, ``Linewidth reduction by feedforward locking a laser diode to a femtosecond comb line,'' \emph{IEEE Photonics Journal}, vol.~8, no.~5, pp. 1--7, 2016.

\bibitem{gatti2012analysis}
D.~Gatti, T.~Sala, A.~Gambetta, N.~Coluccelli, G.~N. Conti, G.~Galzerano, P.~Laporta, and M.~Marangoni, ``Analysis of the feed-forward method for the referencing of a cw laser to a frequency comb,'' \emph{Optics express}, vol.~20, no.~22, pp. 24\,880--24\,885, 2012.

\bibitem{sala2012wide}
T.~Sala, D.~Gatti, A.~Gambetta, N.~Coluccelli, G.~Galzerano, P.~Laporta, and M.~Marangoni, ``Wide-bandwidth phase lock between a cw laser and a frequency comb based on a feed-forward configuration,'' \emph{Optics Letters}, vol.~37, no.~13, pp. 2592--2594, 2012.

\bibitem{fang2015coherence}
S.~Fang, Y.-Y. Jiang, H.-Q. Chen, Y.~Yao, Z.-Y. Bi, and L.-S. Ma, ``Coherence transfer from 1064 nm to 578 nm using an optically referenced frequency comb,'' \emph{Chinese Physics B}, vol.~24, no.~7, p. 074202, 2015.

\bibitem{mcferran2018laser}
J.~McFerran, ``Laser stabilization with a frequency-to-voltage chip for narrow-line laser cooling,'' \emph{Optics Letters}, vol.~43, no.~7, pp. 1475--1478, 2018.

\bibitem{zhang2022frequency}
S.~Zhang, H.~Qiao, D.~Ai, M.~Zhou, and X.~Xu, ``Frequency stabilization of multiple wavelength lasers based on a broadband spectrum,'' \emph{Laser Physics Letters}, vol.~19, no.~9, p. 095701, 2022.

\bibitem{argence2015quantum}
B.~Argence, B.~Chanteau, O.~Lopez, D.~Nicolodi, M.~Abgrall, C.~Chardonnet, C.~Daussy, B.~Darqui{\'e}, Y.~Le~Coq, and A.~Amy-Klein, ``Quantum cascade laser frequency stabilization at the sub-hz level,'' \emph{Nature Photonics}, vol.~9, no.~7, pp. 456--460, 2015.

\bibitem{yasui2021precise}
S.~Yasui, M.~Hiraishi, A.~Ishizawa, H.~Omi, R.~Kaji, S.~Adachi, and T.~Tawara, ``Precise spectroscopy of 167 er: Y 2 sio 5 based on laser frequency stabilization using a fiber laser comb,'' \emph{Optics Express}, vol.~29, no.~17, pp. 27\,137--27\,148, 2021.

\bibitem{miyashita2021offset}
T.~Miyashita, T.~Kondo, K.~Ikeda, K.~Yoshii, F.-L. Hong, and T.~Horikiri, ``Offset-locking-based frequency stabilization of external cavity diode lasers for long-distance quantum communication,'' \emph{Japanese Journal of Applied Physics}, vol.~60, no.~12, p. 122001, 2021.

\bibitem{Zhou:21}
\BIBentryALTinterwordspacing
P.~Zhou, W.~Sun, S.~Liang, S.~Chen, Z.~Zhou, Y.~Huang, H.~Guan, and K.~Gao, ``Digital long-term laser frequency stabilization with an optical frequency comb,'' \emph{Appl. Opt.}, vol.~60, no.~21, pp. 6097--6102, Jul 2021. [Online]. Available: \url{https://opg.optica.org/ao/abstract.cfm?URI=ao-60-21-6097}
\BIBentrySTDinterwordspacing

\bibitem{9511311}
X.~Chen, Q.~Liu, Y.~Wang, F.~Meng, and B.~Luo, ``A high-precision offset frequency locking technique with delay line reference and aom-based compensation,'' \emph{IEEE Photonics Journal}, vol.~13, no.~4, pp. 1--6, 2021.

\end{thebibliography}

\end{document}